\documentclass[a4paper, noarxiv, onecolumn]{quantumarticle}
\usepackage[numbers,sort&compress]{natbib}
\usepackage[utf8]{inputenc}
\usepackage[english]{babel}
\usepackage[title]{appendix}
\usepackage[T1]{fontenc}

\usepackage{etoolbox}
\usepackage{amsmath}
\usepackage{amsfonts}
\usepackage{pgfplots}
\usepackage[dvipsnames]{xcolor}
\usepackage{mathrsfs}
\usepackage{lipsum}
\usepackage{mdframed}
\usepackage{booktabs}
\usepackage{hyperref}
\usepackage{longtable}
\usepackage{color}

\pgfplotsset{compat=1.15}
\usepgfplotslibrary{fillbetween}

\usepackage{tikz}
\usetikzlibrary{quantikz2}

\newtheorem{definition}{Definition}
\newtheorem{proposition}{Proposition}
\newtheorem{theorem}{Theorem}

\newtheorem{lemma}{Lemma}

\begin{document}
	\title{Efficient block-encodings require structure}
	\author{Parker Kuklinski}
	\email{Parker.Kuklinski@ll.mit.edu}
	\author{Benjamin Rempfer}
	\email{Benjamin.Rempfer@ll.mit.edu}
    \author{Justin Elenewski}
	\author{Kevin Obenland}
	\affiliation{MIT Lincoln Laboratory}
	\maketitle
	
    \begin{abstract}
Block-encodings are ubiquitous in quantum computing as a way to represent data within a unitary operator. While several unstructured methods are applicable to arbitrary data, these techniques are burdened by hidden costs and poor accuracy. In this paper, we demonstrate that, even for a small $6$-qubit encoding, these structure-agnostic techniques require wildly intractable resources. We compare these resources with an encoding method which respects a mathematical representation of the block, leading to the conclusion that unstructured encoding mehtods should only be used in the most extenuating circumstances. This finding runs contrary to existing literature on quantum algorithms which often employ structure-agnostic methods.
\end{abstract}
	
	\tableofcontents
	\newpage
    
    \section{Introduction}

Block-encoding is a critical piece of the modern quantum algorithm landscape \cite{clader22}. A block-encoding circuit represents data within a unitary operator, providing quantum access to non-unitary operations. A variety of algorithms take advantage of this concept including quantum phase estimation \cite{morisaki23}, quantum signal processing \cite{lloyd21}, quantum differential equation solvers \cite{wan21}, and quantum option pricing algorithms \cite{stamatopoulos24}. These encodings are particularly critical when manipulating matrix data on a quantum computer.

Despite their prevalent nature, the literature on how to \emph{actually} create a block-encoding is underdeveloped. An often referenced technique is the oracle access model which assumes one has access to two sparse oracles that yield the nonzero entries of the desired block \cite{camps24}; this is typically cited without elucidating how one constructs these oracles for the specific task at hand. Other block-encoding methods have been developed such as FABLE (Fast approximate block-encoding) \cite{fable22}, block-encoding techniques for hierarchical matrices \cite{nguyen22}, quantum sampling algorithms \cite{lemieux24}, block-encodings for the class of pseudodifferential operators \cite{fang23}, block-encodings for linear systems with displacement structures \cite{wan21_1}, QRAM and QROM strategies such as bucket-brigade \cite{clader22} and so forth. These techniques attempt to capture block-encoding circuits for large swathes of possible matrices. It is often assumed that an arbitrary matrix could be handled using one or more of these constructions.

However, the costs associated with these methods are often insurmountable, requiring exponentially many gates (often they are arbitrary rotation gates with large $T$-cost) or an extraordinary number of qubits. This is further exacerbated by the fact that block-encodings are only a single component of whatever algorithm they are servicing; a block-encoding circuit will likely need to be called several times thus multiplying the resources. Furthermore, these bloated generic techniques are frequently invoked in situations which are not appropriate, such as when one needs to block-encode a matrix with a well-defined mathematical representation; typically a generic technique will be used which operates blind to any simplifying structure Alternatively, a generic oracle access model will be invoked under the assumption that the corresponding oracles are \emph{probably} simple to construct if one can write down a mathematical formula for the matrix.

In this contribution, we demonstrate the insidious costs of generic block-encoding methods. This is illustrated explicitly for a simple $6$-qubit matrix arising out of a computational fluid dynamics problem. Even in this small case a block-encoding method which respects a linear algebraic representation of the matrix will perform significantly better than any generic circuit construction. We benchmark these circuits in terms of $T$-cost as the production and consumption of $T$-gates is the rate-limiting temporal resource in fault-tolerant quantum computing. In this way, we advocate to explore bespoke block-encoding circuits in any situation that doesn't involve matrices of raw data.

The rest of our paper is structured as follows: section \ref{two} provides the basic principles of block-encoding and defines the particular $6$-qubit CFD matrix we will block-encode, section \ref{three} gives resource estimates for popular methods of implementing the unstructured oracle encoding circuit as well as our bespoke algebraic encodings and section \ref{four} highlights average-case comparisons of each method. We conclude with a reflection on why these unstructured techniques fail to deliver reasonable costs despite the small size of $F_1$.
    \section{Definitions}
\label{two}
\subsection{Block-Encoding}

We first specify a definition for block-encoding. In the context of this work, we aim to load a matrix of data into a unitary representation $U$ that is compatible with quantum circuits:
\begin{equation}
\label{blockA}
    U=\begin{pmatrix} A & \cdot \\ 
    \cdot & \cdot\end{pmatrix}.
\end{equation}
When we apply a block-encoding circuit $U$ of matrix $A$ to an $a+n$-qubit state $|0\rangle ^{\otimes a}|\psi\rangle$, the result is
\begin{equation}
    U|0\rangle ^{\otimes a}|\psi\rangle =\alpha|0\rangle ^{\otimes a}\left(\tilde{A}|\psi\rangle\right) +\sqrt{1-\left(|\alpha |\lVert\tilde{A}|\psi\rangle\rVert\right)^2}|\phi _g\rangle
\end{equation}
where $\tilde{A}$ is an approximation of $A$, $\alpha$ is the \emph{subnormalization factor} (a smaller subnormalization means one must run more iterations of amplitude amplification \cite{brassard02} to recover the desired state), and $|\phi _g\rangle$ is an $a+n$-qubit garbage state. Unless otherwise stated, we take $\lVert\cdot\rVert$ to be the $L^2$ norm either as an operator norm or a vector norm. The appeal of block-encoding is that it allows us to apply non-unitary linear operators to quantum states. We formally state the definition:

\begin{definition}
Let $a_c,a_p,m,n\in\mathbb{N}$ and $m=a_c+a_p+n$ where $m$ is the total number of qubits, $n$ is the number of data qubits, $a_c$ is the number of clean ancilla (ancilla beginning and ending in the $|0\rangle$ state), and $a_p$ is the number of persistant ancilla ending in a garbage state. We say that an $m$-qubit unitary operator $U$ is a $(\alpha ,(a_c,a_p),\epsilon )$-\emph{block-encoding} of an n-qubit operator $A$ with $\lVert A\rVert =1$ (not necessarily unitary) if $\lVert A-\tilde{A}\rVert <\epsilon$ where
\begin{equation}
    \tilde{A}=\frac{\left(\langle 0|^{\otimes a}\otimes I_n\right) U\left( |0\rangle ^{\otimes a}\otimes I_n\right)}{\alpha}
\end{equation}
and $\alpha =\left\lVert\left(\langle 0|^{\otimes a}\otimes I_n\right) U\left( |0\rangle ^{\otimes a}\otimes I_n\right)\right\rVert$.
\end{definition}

We will ``normalize'' each matrix $A$ such that $\lVert A\rVert=1$; to avoid a situation where the error is reported as small simply by virtue of the subnormalization being small. This is a reasonable choice since the $L^2$ norm of any sub-block of a unitary matrix is bounded above by 1, as the following lemma demonstrates:

\begin{lemma}
Let $U=\begin{pmatrix} A & B \\ C & D\end{pmatrix}$ be an arbitrary block matrix. Then $\lVert A\rVert\le \lVert U\rVert$.
\end{lemma}

{\bf Proof:} Recall that any matrix norm can be defined as
\begin{equation}
    \lVert A\rVert =\max _{\lVert x\rVert =1}\lVert Ax\rVert.
\end{equation}
Let $x$ be the particular unit vector which maximizes $\lVert Ax\rVert$ such that $\lVert A\rVert =\lVert Ax\rVert$ (possible because $A$ is finite dimensional). If we define $\tilde{x}$ as $x$ concatenated with zeros to fit the dimension of $U$, then $\tilde{x}$ is a unit vector and we have
\begin{equation}
    \lVert A\rVert =\lVert Ax\rVert \le\lVert U\tilde{x}\rVert\le\lVert U\rVert
\end{equation}
thus completing the proof. $\hfill\Box$

Using this lemma, a large subnormalization close to this upper bound of $1$ indicates our block-encoding circuit is efficient in this metric. Despite an effort to normalize our encodings in order to keep error consistent, the mathematics of calculating error will often be easier in an unnormalized setting. To compensate for this, we will leverage the following lemma to pass from an unnormalized error bound to a normalized one.

\begin{lemma}
If $\lVert A-\tilde{A}\rVert<\epsilon$ and $d>2$, then $\epsilon <\lVert A\rVert (1-2/d)$ implies
\begin{equation}
    \left\lVert\frac{A}{\lVert A\rVert}-\frac{\tilde{A}}{\lVert \tilde{A}\rVert}\right\rVert <\frac{d\epsilon}{\lVert A\rVert.}
\end{equation}
\label{lemmaBOUND}
\end{lemma}

{\bf Proof:} Using standard triangle inequality comparisons, we can prove
\begin{equation}
\label{bound1}
    \left\lVert\frac{A}{\lVert A\rVert}-\frac{\tilde{A}}{\lVert \tilde{A}\rVert}\right\rVert <\frac{\epsilon}{\lVert A\rVert} +(\epsilon +\lVert A\rVert)\left\lVert\frac{1}{\lVert A\rVert}-\frac{1}{\lVert\tilde{A}\rVert}\right\rVert.
\end{equation}
Solving the algebra gives us that for $c>1$, $\epsilon <\lVert A\rVert\left( 1-1/c\right)$ we have
\begin{equation}
\label{bound2}
    \left\lVert\frac{1}{\lVert A\rVert}-\frac{1}{\lVert\tilde{A}\rVert}\right\rVert <\frac{c\epsilon}{\lVert A\rVert ^2}.
\end{equation}
Substituting equation (\ref{bound2}) into equation (\ref{bound1}) and further imposing that $\epsilon <\lVert A\rVert (d-(1+c))/c$ gives
\begin{equation}
\label{bound3}
    \left\lVert\frac{A}{\lVert A\rVert}-\frac{\tilde{A}}{\lVert \tilde{A}\rVert}\right\rVert <\frac{d\epsilon}{\lVert A\rVert}.
\end{equation}
Equations (\ref{bound2}) and (\ref{bound3}) are equal when $d=2c$, and a substitution proves the result. $\hfill\Box$ 

In the cases we will consider, $\epsilon$ is orders of magnitude less than $\lVert A\rVert$ so it suffices to consider the bound $d=2$. To evaluate the block-encoding techniques we will count the number of $T$-gates \cite{brylinski02} required to implement the circuit $U$ such that $A$ is encoded to accuracy $\epsilon$, call this $T(\epsilon )$. Since the number of amplitude amplification iterations required for the final measurement scales as $O(1/\alpha )$ \cite{brassard02}, we will consider our figure of comparison to be $T(\epsilon )/\alpha$ as a proxy for the number of $T$-gates needed to successfully produce a quantum state operated on by the block.
    \subsection{CFD Matrix}

Our discussion will be driven by the block encoding of a $6$-qubit matrix called $F_1$. This matrix arises from a computational fluid dynamics problem and represents a linear approximation to a nonlinear differential equation. Penuel et. al. details the relevance of such an algorithm in the context of ship hull design \cite{penuel24}; in this case the matrix represents the motion of a discretized fluid to nearest-neighbor cells in a $3\times 3\times 3$ cube. 

Let $c=[c_{x},c_{y},c_{z}]\in\{ -1,0,1\} ^{64\times 3}$ be a matrix containing in its rows all 27 vectors corresponding to lattice points in a $3\times 3\times 3$ unit cube centered at the origin. Then extend $c$ with an additional 37 rows of arbitrary elements (padding with an extra 32 rows to fit the matrix to 6 qubits will help with the following algebraic encoding). Let $c_i\in\{ -1,0,1\} ^3$ be the $i^\text{th}$ vector in such a list. Let $w_i=(1/4)^{|c_i|}$ be a weighting function which essentially `penalizes' each item $i$ by a multiplicative factor of $1/4$ for each nonzero element $c_i$ contains.

Penuel et. al. \cite{penuel24} initially defines $F_1$ by element in the following way (up to constant):

\begin{align}
    (F_1)_{ii}&=-1+w_i+3w_i(c_i\cdot c_i) \\
    (F_1)_{ij}&=w_i(1+3(c_i\cdot c_j)).
\end{align}

\begin{figure}
    \centering
    \includegraphics[scale=0.15]{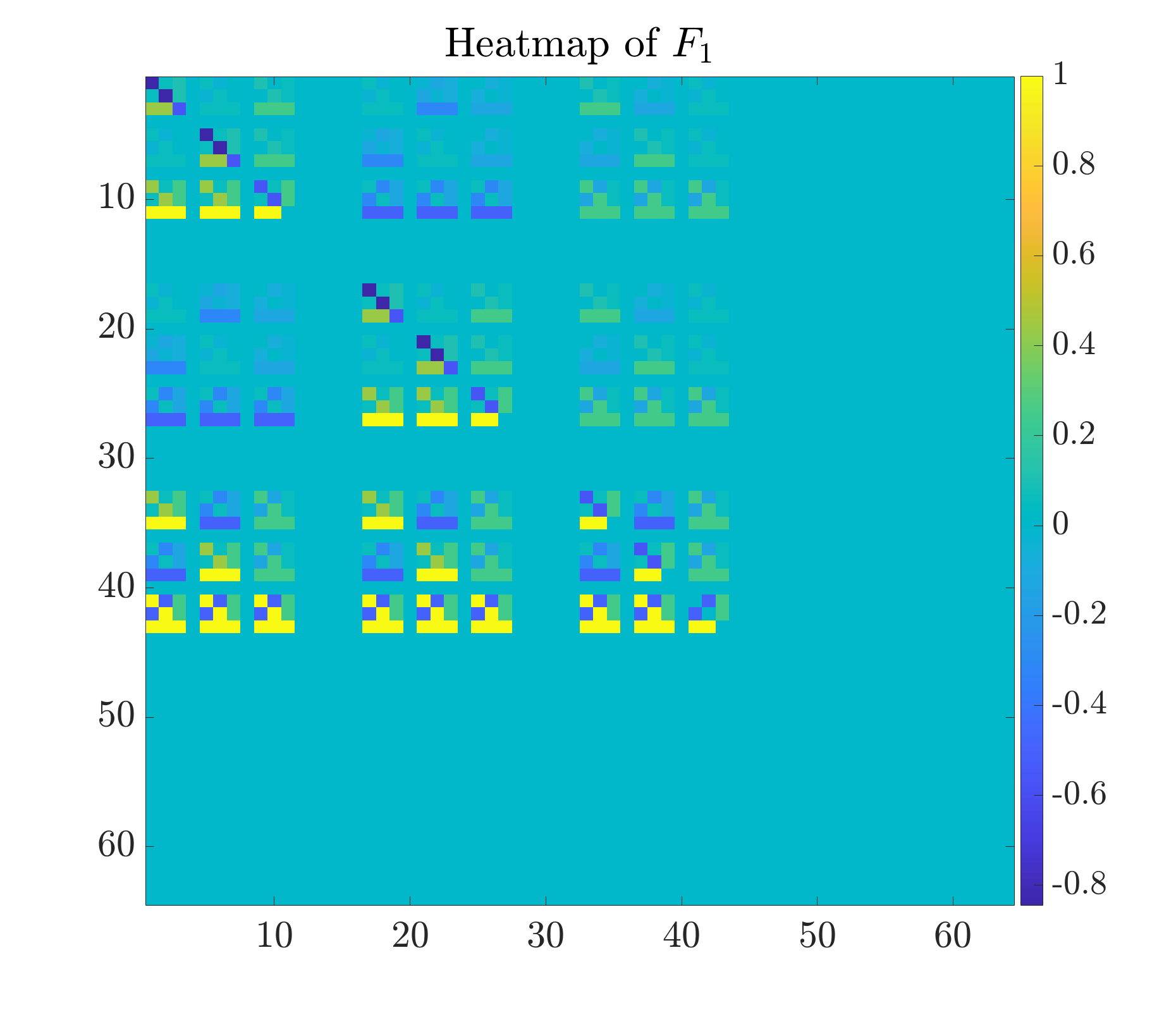}
    \caption{Heatmap of the matrix $F_1$ as defined in equation (\ref{F1}).}
    \label{f1_mat}
\end{figure}

Here $i$ is taken to be the index of a 27 element list. $F_1$ has 722 nonzero elements and 14 unique elements as can be ascertained from Figure \ref{f1_mat}.

In contrast to the above equations we present a more convenient matrix algebra representation (thus facilitating a circuit construction). Let $C=cc^T$ such that $(C)_{ij}=(c_i\cdot c_j)$. Let $W=\text{diag}(w)$ be a diagonal matrix with $(W)_{ii}=w_i$. Let ${\bf 1}_{6}$ be a 6-qubit matrix filled with ones and let $I_6$ be the 6-qubit identity matrix. Finally, let $P$ be a diagonal projection off of the base-4-padded elements such that $(P)_{ii}=1$ if $c_i$ corresponds to one of the lattice points in the cube and $(P)_{ii}=0$ otherwise. This allows us to write

\begin{equation}
\label{F1}
    F_1=P(W({\bf 1}_6+3C)-I_6)P
\end{equation}

By a clever ordering, we can also represent $C$ in terms of tensor products. Let $x=[1,-1,0]'$ and $o=[1,1,1]'$. Then we can see that choosing

\begin{align}
\label{tensor0}
    c_x&=o\otimes o\otimes x \\
\label{tensor1}
    c_y&=o\otimes x\otimes o \\
\label{tensor2}
    c_z&=x\otimes o\otimes o
\end{align}

corresponds to the 27 unique vectors in a $3\times 3\times 3$ unit cube. If we append both $x$ and $o$ with an additional quantity (these will be 0 and 1 respectively), this recovers the $64\times 3$ $c$ matrix. Further, if we let $w=\text{diag}\left( [\frac{1}{4},\frac{1}{4},1,1]\right)$, then we can suggestively write $W=w\otimes w\otimes w$.
    \subsection{Unstructured Oracle Encodings}
\label{unstructuredAPP}

Before detailing how we plan to encode the $F_1$ matrix from equation (\ref{F1}) we must outline the unstructured oracle model for block-encoding an arbitrary matrix. If one has access to an oracle which encodes all of the entries of the $2^n\times 2^n$ block on the diagonal of a $2^{2n}\times 2^{2n}$ matrix, then the following circuit rearranges those entries into the desired block:

\begin{figure}
\[
\begin{quantikz}
\lstick{$|0\rangle$} & & \gate[3]{O_{H^{\otimes n}AH^{\otimes n}}} & & & \meter{} \\
\lstick{$|0\rangle^{\otimes n}$} & \gate{H^{\otimes n}} & & \swap{1} & \gate{H^{\otimes n}} & \meter{} \\
\lstick{$|\psi\rangle$} & & & \targX{} & & & \rstick{$\frac{A|\psi\rangle}{\lVert A|\psi\rangle\rVert}$}
\end{quantikz}
\]
\caption{Unstructured oracle block-encoding circuit.}
\label{fig1}
\end{figure}
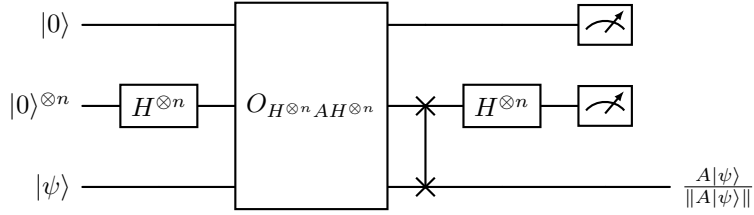

\begin{theorem}
Let $A=(a_{ij})$ be a real $2^n\times 2^n$ matrix with $\lVert A\rVert =1$ where $|A|$ is the maximum absolute value element, and let $O_A$ be an oracle operator defined on a $(2n+1)$--qubit Hilbert space:
\begin{equation}
    O_A|0\rangle |i\rangle |j\rangle =\left( \frac{a_{ij}}{|A|}|0\rangle +\sqrt{1-\frac{|a_{ij}|^2}{|A|^2}}|1\rangle\right) |i\rangle |j\rangle.
\end{equation}
Then the operator $U_A$ described in equation (\ref{UA}) (depicted in Figure \ref{fig1}) is a $\left( 1/(|A|2^{n}),(0,n+1),0\right)$-block-encoding of $A$
\begin{equation}
\label{UA}
U_A=\left( I_1\otimes H^{\otimes n}\otimes I_n\right)\left( I_1\otimes\text{SWAP}\right) O_A\left( I_1\otimes H^{\otimes n}\otimes I_n\right).
\end{equation}
\end{theorem}
{\bf Proof:} See \cite{fable22}. $\hfill\Box$

\begin{figure}
\centering
\includegraphics[scale=0.17]{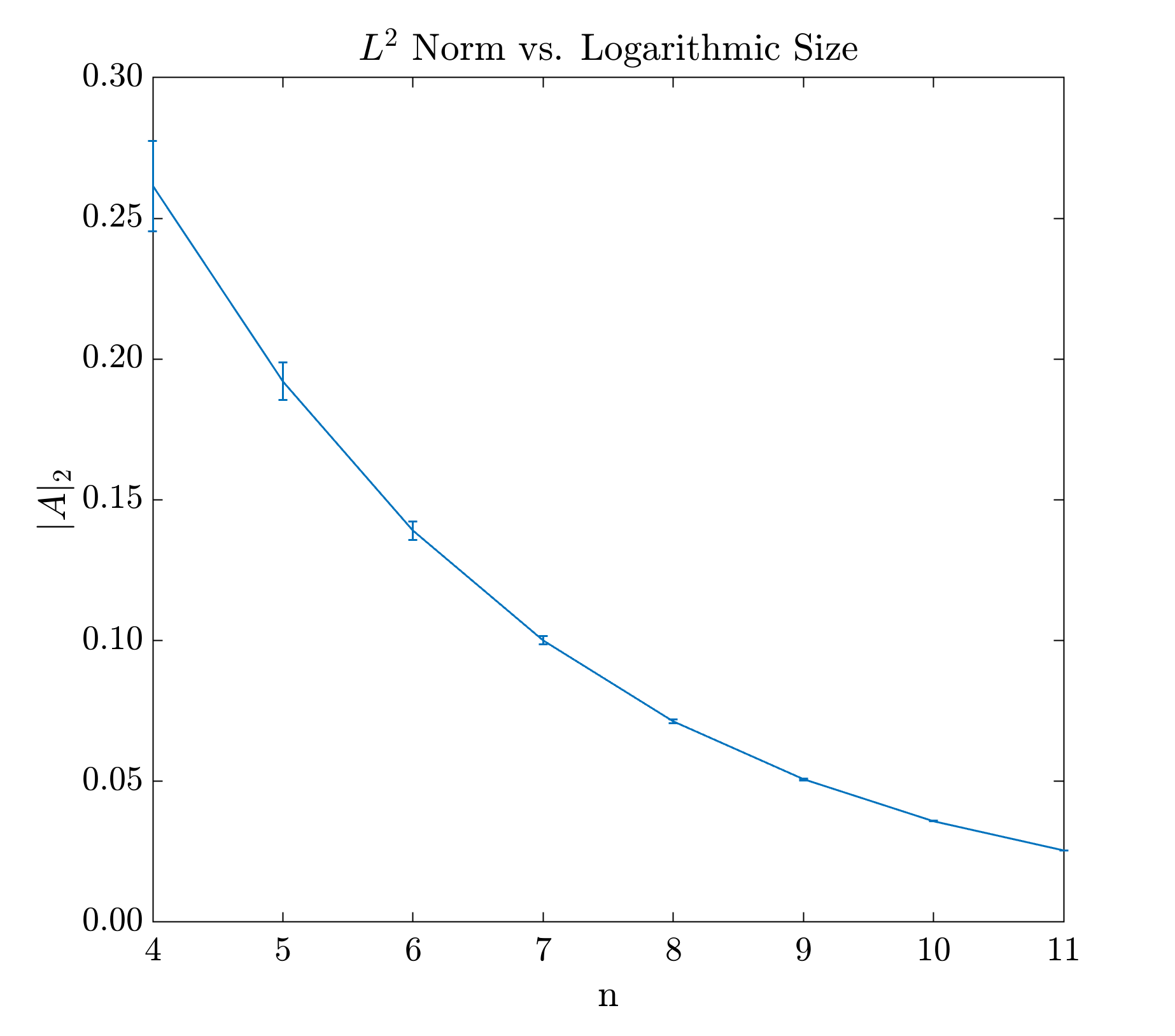}
\caption{Plot of $L^2$ norms of randomly simulated matrices size $2^n\times 2^n$. In each matrix $A=(a_{ij})$, where $a_{ij}$ is a uniformly distributed random variable from $[-\frac{1}{2^n},\frac{1}{2^n}]$. Thus $A$ is a random matrix encoded by an unstructured oracle circuit and its $L^2$ norm is the subnormalization factor. Error bars indicate smallest and largest $L^2$ norms of $100$ simulated matrices.}
\label{fig2}
\end{figure}

In this construction the matrix being encoded in the block is a scaled copy of $A$ with element of maximum absolute value $1/2^n$; this is due to the diagonal elements of $O_A$ being necessarily bounded by absolute value 1 and the factor of $1/2^n$ comes from the Hadamard gates. While this may seem deleterious for the subnormalization (e.g. if $A$ was the identity matrix the subnormalization factor would be $1/2^n$), for unstructured matrices the subnormalization is on average approximately $1.13/2^{n/2}$ as seen in Figure \ref{fig2}. While this subnormalization is not so daunting, it is still exponential in size which is far from ideal.

The unstructured oracle model is agnostic to the specific circuit implementing $O_A$; if an efficient implementation exists then one only needs to contend with the exponentially small subnormalization. In practice, it is not so simple to conceive of an efficient oracle in this framework even for structured matrices. Generally, oracle constructions require one rotation per matrix element, or in some cases one rotation per unique matrix element. If there are exponentially many elements in a block, or if the small number of unique elements are arranged in an indecipherable pattern, these costs will quickly get out of hand. We document three common oracle constructions below.

\subsubsection{Unary Iteration}

We first illustrate a standard unary iteration procedure \cite{babbush18} for constructing the oracle $O_A$. For each matrix element $a_{ij}$, we implement a multi-qubit controlled rotation gate with controls on the $|i\rangle |j\rangle$ register and target on the top register that encodes $a_{ij}/|A|$ on the $(2^ni+j)^\text{th}$ diagonal element of the total unitary. Each of these multi-controlled gates can be expanded into a pyramid of Toffoli gates implemented with $2n-1$ clean ancilla and costing $8n-4$ $T$-gates \cite{babbush18}; if there are $2^{2n}$ distinct elements then the total $T$-cost would be the cost of $2^{2n}$ controlled rotations to accuracy plus $(8n-4)2^n$ $T$-gates for the multi-controls.

\begin{figure}
\[
\resizebox{\columnwidth}{!}{
\begin{quantikz}
& \gate{\theta_0} & \gate{\theta_1} & \gate{\theta_2} & \gate{\theta_3} & \gate{\theta_4} & \gate{\theta_5} & \gate{\theta_6} & \gate{\theta_7} & \gate{\theta_8} & \gate{\theta_9} & \gate{\theta_{10}} & \gate{\theta_{11}} & \gate{\theta_{12}} & \gate{\theta_{13}} & \gate{\theta_{14}} & \gate{\theta_{15}} & \\
& \ctrl[open]{-1} & \ctrl[open]{-1} & \ctrl[open]{-1} & \ctrl[open]{-1} & \ctrl[open]{-1} & \ctrl[open]{-1} & \ctrl[open]{-1} & \ctrl[open]{-1} & \ctrl{-1} & \ctrl{-1} & \ctrl{-1} & \ctrl{-1} & \ctrl{-1} & \ctrl{-1} & \ctrl{-1} & \ctrl{-1} & \\
& \ctrl[open]{-1} & \ctrl[open]{-1} & \ctrl[open]{-1} & \ctrl[open]{-1} & \ctrl{-1} & \ctrl{-1} & \ctrl{-1} & \ctrl{-1} & \ctrl[open]{-1} & \ctrl[open]{-1} & \ctrl[open]{-1} & \ctrl[open]{-1} & \ctrl{-1} & \ctrl{-1} & \ctrl{-1} & \ctrl{-1} & \\
& \ctrl[open]{-1} & \ctrl[open]{-1} & \ctrl{-1} & \ctrl{-1} & \ctrl[open]{-1} & \ctrl[open]{-1} & \ctrl{-1} & \ctrl{-1} & \ctrl[open]{-1} & \ctrl[open]{-1} & \ctrl{-1} & \ctrl{-1} & \ctrl[open]{-1} & \ctrl[open]{-1} & \ctrl{-1} & \ctrl{-1} & \\
& \ctrl[open]{-1} & \ctrl{-1} & \ctrl[open]{-1} & \ctrl{-1} & \ctrl[open]{-1} & \ctrl{-1} & \ctrl[open]{-1} & \ctrl{-1} & \ctrl[open]{-1} & \ctrl{-1} & \ctrl[open]{-1} & \ctrl{-1} & \ctrl[open]{-1} & \ctrl{-1} & \ctrl[open]{-1} & \ctrl{-1} &
\end{quantikz}}
\]
\caption{Na\"{i}ve implementation of an unstructured block encoding for a $4\times4$ matrix. Each $\theta _{ij}$ refers to a rotation gate $R_y(\theta _{ij})$ which encodes the $ij$ matrix element $\sin (\theta _{ij})/2^n$.}
\label{naive}
\end{figure}
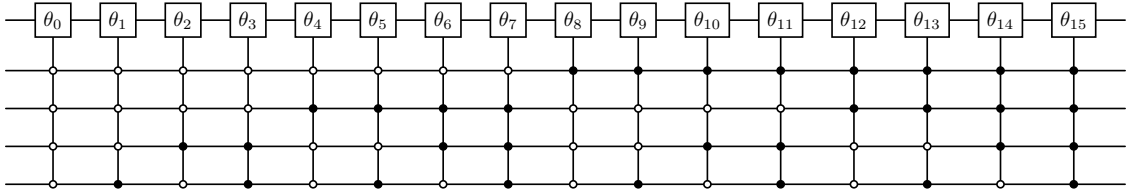

Fortunately, we can cut down on these costs in a few ways. By using the unary iteration scheme from \cite{babbush18}, the multi-control scheme costs approximately $2^{2n+2}$ $T$-gates. Further, if there is sparsity in the matrix or many elements with absolute value equal to $|A|$, we can restructure the oracle (via Clifford gates) to replace approximate multi-controlled rotations with multi-controlled $X$ gates (in the case of elements equaling $|A|$) or outright eliminating rotations in the case of vanishing matrix elements. Doing this will help alleviate the accuracy requirements on the remaining rotation gates; if $A$ only contains elements $a_{ij}\in\{ -|A|,0,|A|\}$ then this oracle can be constructed with $\epsilon =0$. 

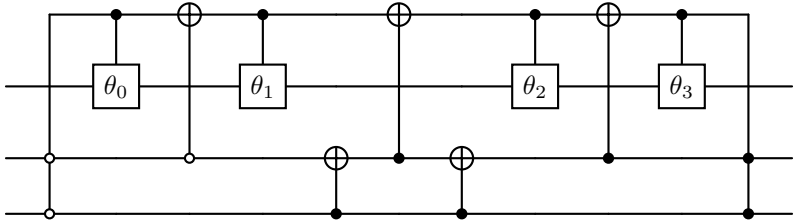
\begin{figure}
\[
\begin{quantikz}
& \wireoverride{n} & \ctrl{1} & \targ{} & \ctrl{1} & & \targ{} & & \ctrl{1} & \targ{} & \ctrl{1} & \\
& & \gate{\theta_0} & & \gate{\theta_1} & & & & \gate{\theta_2} & & \gate{\theta_3} & & \\
& \ctrl[open]{-2} & & \ctrl[open]{-2} & & \targ{} & \ctrl{-2} & \targ{} & & \ctrl{-2} & & \ctrl{-2} & \\
& \ctrl[open]{-1} & & & & \ctrl{-1} & & \ctrl{-1} & & & & \ctrl{-1} &
\end{quantikz}
\]
\caption{Unary iteration implementation of an unstructured block-encoding for a $2\times2$ matrix. Rather than expanding each multi-control gate via Toffolis (would coust $4$ Toffoli pairs, one for each gate), the present construction uses just one Toffoli pair. The top qubit is clean (i.e. starts and ends in $|0\rangle$).}
\label{unary}
\end{figure}

In spite of these optimizations, there will still be an exponential number of resources required to implement this oracle (unless $A$ is unusually sparse with $O(\text{poly}(n))$ nonzero elements total).

\subsubsection{QROM}

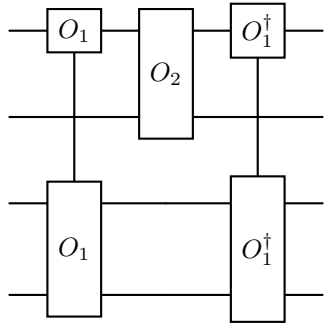
\begin{figure}
\[
\begin{quantikz}
& \gate{O_1} & \gate[2]{O_2} & \gate{O_1^\dag} & \\
& & & & \\
& \gate[2]{O_1}\wire[u][2]{q} & & \gate[2]{O_1^\dag}\wire[u][2]{q} & \\
& & & & 
\end{quantikz}
\]
\caption{Example QROM oracle implementation. $O_1$ acts as $O_1|0\rangle |0\rangle |i\rangle |j\rangle =|u_{ij}\rangle |0\rangle |i\rangle |j\rangle$, where $u_{ij}$ denotes the index of the unique element. This can be implemented with only as many multi-controlled $X$ gates as there are nonzero elements in the matrix. $O_2$ meanwhile maps $O_2|u_{ij}\rangle |0\rangle =|u_{n}\rangle\left( R_y(\theta _{n})|0\rangle \right)$, implementing the desired rotation on the second qubit corresponding to the index contained in the first. $O_2$ thus only contains as many controlled rotations as there are unique elements in the matrix. The application of $O_1^{\dagger}$ uncomputes the top register.}
\label{QROM}
\end{figure}

If many of the rotation gates are the same, we can further collate resources and use same number of rotation gates as there are unique elements in the matrix. This can be done by encoding the unique rotations into a smaller oracle and then accessing this for the individual matrix elements with unary iteration composed only of multi-controlled $X$ gates as in Figure \ref{QROM}. While this clearly cuts down on the number of rotation gates, the cost of implementing the multi-control scheme still remains.

\subsubsection{S-FABLE}

To eliminate the multi-control sequence from the unary iteration scheme, Camps and Van Beeuman \cite{fable22} developed the ``Fast Approximate Block-Encoding'' method (FABLE). Rather than implementing a sequence of multi-controlled rotations, this technique uses rotation identities to implement an equivalent circuit with only $CX$ gates and uncontrolled rotations on a new set of rotation angles; the only $T$-gates arise from these rotations. In particular, if $\text{vec}(\theta)$ is the vector of original rotation angles in Figure \ref{naive}, the new rotation angles $\text{vec}(\hat{\theta})$ can be expressed as
\begin{equation}
    \text{vec}(\hat{\theta})=\frac{1}{N}(H^{\otimes 2n}P_G)^{-1}\text{vec}(\theta )
\end{equation}
where $H$ is the Hadamard matrix and $P_G$ is a Gray code permutation \cite{doran07}.

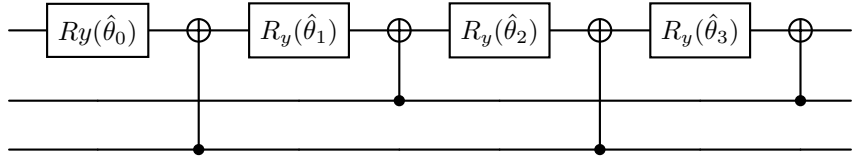
\begin{figure}
\[
\begin{quantikz}
& \gate{Ry(\hat{\theta}_0)} & \targ{} & \gate{R_y(\hat{\theta}_1)} & \targ{} & \gate{R_y(\hat{\theta}_2)} & \targ{} & \gate{R_y(\hat{\theta}_3)} & \targ{} & \\
& & & & \ctrl{-1} & & & & \ctrl{-1} & \\
& & \ctrl{-2} & & & & \ctrl{-2} & & & 
\end{quantikz}
\]
\caption{FABLE implementation for an oracle corresponding to a $2\times 2$ matrix block-encoding. Since there are no multi-controlled gates, the only $T$-gates in this circuit arise from the uncontrolled rotation gates.}
\label{FABLE}
\end{figure}

While this method seems promising in that it eliminates all multi-controlled gates, there are several drawbacks which have not been addressed, notwithstanding the immense classical pre-processing. First and most importantly, the angles of the transformed uncontrolled rotations are exponentially smaller than the original angles. This means approximations to these new rotation gates must be exponentially more accurate to achieve the same total error, leading to a multiplicative gate cost of $O(n)$. Furthermore, there are no guarantees that the transformed angle set captures the same simplifying structure from the original set. It was mentioned in Kuklinski and Rempfer \cite{kuklinski24} that this angle transformation will eliminate sparsity (ostensibly the `fast' part of FABLE), but it may also explode the number of unique rotations. From these observations, only certain matrices may be highly compressible under the FABLE transformation and it is difficult \emph{a priori} to determine which matrices perform well. From \cite{kuklinski24} it seems that many structured matrices actually perform worse than the average case.

To this end, in this paper we use a modification from Kuklinski and Rempfer \cite{kuklinski24} called S-FABLE (sparse-FABLE). Since it was observed that matrices sparse in the Hadamard-Walsh domain have compressible oracles under FABLE (i.e. $H^{\otimes n}AH^{\otimes n}$ is sparse), we can more efficiently block-encode a sparse matrix $A$ by using FABLE to block-encode $H^{\otimes n}AH^{\otimes n}$ (since $H^{\otimes n}\left( H^{\otimes n}AH^{\otimes n}\right) H^{\otimes n}=A$ is sparse) and conjugating this block-encoding on the quantum computer with $H$ gates to recover a block-encoding of $A$. This incurs additional classical costs of computing the Hadamard-Walsh transform of the target block, however in the case of a 6-qubit matrix this is trivial.

\begin{figure}
\[
\begin{quantikz}
\lstick{$|0\rangle$} & & \gate[3]{O_{H^{\otimes n}AH^{\otimes n}}} & & & \meter{} \\
\lstick{$|0\rangle^{\otimes n}$} & \gate[2]{H^{\otimes2n}} & & \swap{1} & \gate[2]{H^{\otimes2n}} & \meter{} \\
\lstick{$|\psi\rangle$} & & & \targX{} & & & \rstick{$\frac{A|\psi\rangle}{\lVert A|\psi\rangle\rVert}$}
\end{quantikz}
\]
\caption{S-FABLE circuit. If $A$ is sparse, a FABLE oracle is used to encode $H^{\otimes n}AH^{\otimes n}$, and an extra pair of Hadamard tensors on the data qubit register conjugates the result to get back $A$. If $A$ is sparse, this will be more efficient than a FABLE circuit.}
\end{figure}
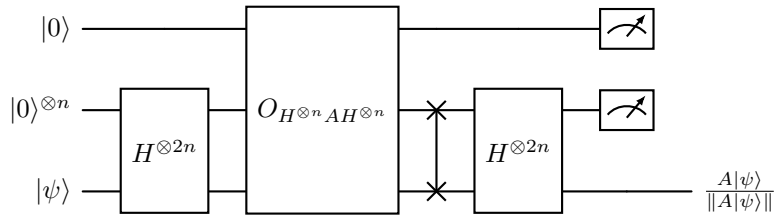
    \section{Resource Estimates}
\label{three}

\begin{figure}
\[
\begin{quantikz}[align equals at=1.5]
& \ctrl{1} & \\
& \gate{R_\alpha(\theta)} &
\end{quantikz}
=\begin{quantikz}[align equals at=1.5]
& & \ctrl{1} & & \ctrl{1} & \\
& \gate{R_\alpha(\theta/2)} & \targ{} & \gate{R_\alpha(-\theta/2)} & \targ{} &
\end{quantikz}
\]\vspace{1cm}\[
\begin{quantikz}[align equals at=1.5]
& \ctrl{1} & \\
& \gate{H} &
\end{quantikz}
=\begin{quantikz}[align equals at=1.5]
& & & & \ctrl{1} & & & \\
& \gate{S} & \gate{H} & \gate{T} & \targ{} & \gate{T^\dag} & \gate{H} & \gate{S^\dag} & 
\end{quantikz}
\]
\caption{Decompositions of commonly occurring two-qubit gates. (\emph{Top}) Decomposition of a controlled-Hadamard gate into Clifford+$T$. (\emph{Bottom}) Decomposition of controlled rotation gate into uncontrolled rotations and $CX$s.}
\label{controlledRotation}
\end{figure}
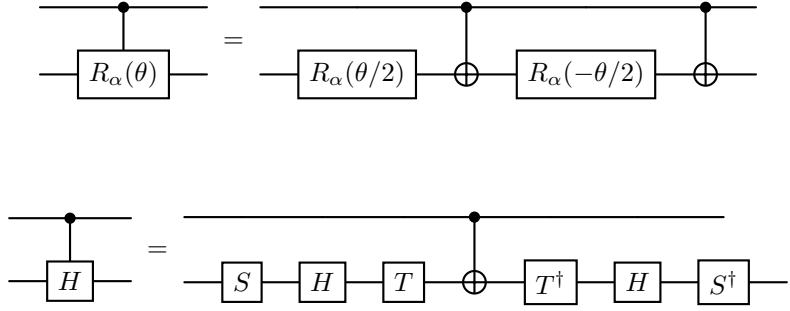

Equipped with these definitions we state the resources required for each of the block-encoding methods. We will use a result from \cite{bocharov15} that an $\epsilon$-approximation of an arbitrary rotation gate costs approximately $1.15\log _2(1/\epsilon )+9.2$ $T$-gates. Using the circuit in Figure \ref{controlledRotation}, this implies that an $\epsilon$-approximate controlled rotation costs $2(1.15\log _2(2/\epsilon )+9.2)=2.3\log _2(1/\epsilon )+20.7$ $T$-gates. We also make use of the fact that a controlled-Hadamard costs two $T$-gates (circuit in Figure \ref{controlledRotation}) and that a pair of Toffoli gates costs four $T$-gates.

\subsection{Unstructured Oracle}

\paragraph{Unary iteration}

We first cost out the unary iteration implementation for the oracle corresponding to $F_1$. Since $F_1$ is not fully dense, we can eliminate multi-control gates corresponding to zero elements. We need the following lemma to determine how many Toffoli pairs can be eliminated between adjacent multi-control gates (as in Figure \ref{unary}).

\begin{lemma}
\label{unaryLEMMA}
Let $a_1a_2...a_n$ be a binary representation for the $n$ qubits of controls on a controlled-$U_1$ operation and let $b_1b_2...b_n$ be a binary representation for the controls on a controlled-$U_2$ operation. If we put these operations in sequence, then if $a_j=b_j$ for $j<k$ but $a_k\ne b_k$ then we can eliminate $k-1$ Toffoli pairs. If $a_1\ne b_1$ we can still eliminate a single Toffoli pair.
\end{lemma}

Now we are ready to present the result:

\begin{theorem}
\label{unaryTHM}
There exists a  unary iteration circuit that is a $(0.0905,(11,7),\epsilon )$ block-encoding of $F_1$ which costs approximately $1490.4\log _2(1/\epsilon )+22368$ T-gates to implement.
\end{theorem}
{\bf Proof:} See Appendix \ref{circuits}.

\paragraph{QROM}

The QROM circuit is structured similarly to the unary iteration circuit except instead of redundant rotations we encode the $14$ unique rotations into a separate $4$-qubit oracle which we call with multi-controlled X gates. We compile this into the following result:

\begin{theorem}
There exists a QROM circuit that is a $(0.0905,(15,7),\epsilon )$ block-encoding of $F_1$ which costs approximately $27.6\log_2(1/\epsilon )+9062$ T-gates to implement.
\label{QROMTHM}
\end{theorem}
{\bf Proof:} See Appendix \ref{circuits} \\

While the constant overhead from the QROM implementation is about $40\%$ of the overhead from the simple unary iteration circuit, the fact that QROM only uses 12 controlled rotations rather than 648 greatly improves cost savings as the accuracy requirement increases.

\paragraph{S-FABLE}

We now document the costs associated with block-encoding $F_1$ using a sparse alternative of FABLE called S-FABLE. In a vanilla implementation of FABLE, many of the rotations in Figure \ref{FABLE} can be eliminated if they are close to zero; this happens when the block-encoded matrix $A$ is approximately sparse in the Hadamard-Walsh domain (i.e. if $H^{\otimes n}AH^{\otimes n}$ is sparse). To block-encode a sparse matrix such as $F_1$ ($\sim 18\%$ sparse), S-FABLE dictates that we use FABLE to block-encode $H^{\otimes n}F_1H^{\otimes n}$ (which is Hadamard-Walsh sparse since $H^{\otimes n}\left( H^{\otimes n}F_1H^{\otimes n}\right) H^{\otimes n}=F_1$ is sparse) and then apply $n$ Hadamards on both sides of the FABLE circuit to recover $F_1$.

\begin{theorem}
\label{FABLE_THM}
There exists an S-FABLE circuit that is a $(0.0903,(0,7),\epsilon )$ block-encoding of $F_1$ which costs approximately
\[ \begin{cases} 
      (771.8\log _2 (1/\epsilon )+483)(1.15\log _2 (1/\epsilon )+21.24) & 7.08\times 10^{-4}<\epsilon \\
      4710.4\log _2(1/\epsilon)+86999 & \epsilon < 7.08\times 10^{-4} 
   \end{cases}
\]
T-gates to implement.
\end{theorem}
{\bf Proof:} See Appendix \ref{circuits}.

This result demonstrates the severe shortcomings of FABLE based methods. What resources are saved by only using uncontrolled rotations and Clifford gates are spent many times over since (1) the rotation angles are exponentially small, (2) `sparsity' in the sense of the ability to eliminate negligible rotation gates only holds up to a rather large threshold accuracy, and (3) there is no ability to exploit repeated elements. 

\subsection{Algebraic Encodings}

After working through resource estimates of block-encoding $F_1$ using unstructured oracle techniques, we now present two much more efficient block-encoding circuits which reflect the mathematics taking place in equation (\ref{F1}). We develop constituent block-encodings for each of the terms in the equation and combine them together using an LCU \cite{childs12}.

\paragraph{Gate-count optimized}

We first tackle encoding the matrix $C$. Using the representation given by equations (\ref{tensor0}), (\ref{tensor1}), and (\ref{tensor2}) it is clear that we need access to both the vector $x$ and the vector $o$ of ones in some capacity. To this end, let $A=[o,x]$ be the $3\times 2$ matrix containing each of these column vectors and consider the following block-encoding.

\begin{figure}
\[
\begin{quantikz}
& \gate{H} & \ctrl{1} & \gate{H} & \\
& & \targ{} & \gate{H} & \\
& & \ctrl{-1} & \gate{H} &
\end{quantikz}
\]
\caption{Gate-optimized block-encoding circuit for $A$}
\label{gateA}
\end{figure}
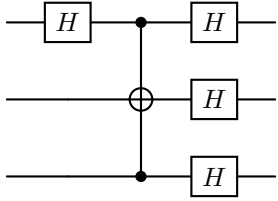

\begin{proposition}
The block-encoding in Figure \ref{gateA} is a $(0.866,(1,1),0)$ block-encoding of $A$ which costs four $T$-gates uncontrolled and sixteen $T$-gates controlled.
\end{proposition}

We include the controlled $T$-count since this block-encoding will eventually be implemented in an LCU. To gain intuition for this circuit construction, consider the following sum:

\begin{equation}
    A=\begin{pmatrix}
1 & 1 & \cdot & \cdot \\
1 & -1 & \cdot & \cdot \\
1 & 0 & \cdot & \cdot \\
\cdot & \cdot & \cdot & \cdot
\end{pmatrix}=
\frac{1}{2}\begin{pmatrix}
1 & 1 & 1 & 1 \\
1 & -1 & 1 & -1 \\
1 & 1 & -1 & -1 \\
1 & -1 & -1 & 1
\end{pmatrix}
+\frac{1}{2}\begin{pmatrix}
1 & 1 & 1 & 1 \\
1 & -1 & 1 & -1 \\
1 & -1 & -1 & 1 \\
1 & 1 & -1 & -1
\end{pmatrix}.
\end{equation}

The first matrix in the sum is simply $H\otimes H$ and the second results from applying a CNOT before $H\otimes H$. Thus, we can encode $A$ by combining these into a simple uniform sum LCU using only a single Hadamard for PREP which importantly does not introduce error to the circuit.

Now that we have a block-encoded $A$, we can recover $c_x,c_y,c_z$ by taking tensor products; namely $A^{\otimes 3}$ contains $c_x$ in its $2^\text{nd}$ column, $c_y$ in its $5^\text{th}$ column, and $c_z$ in its $17^\text{th}$ column. In this way, we can block-encode $c$ very naturally.

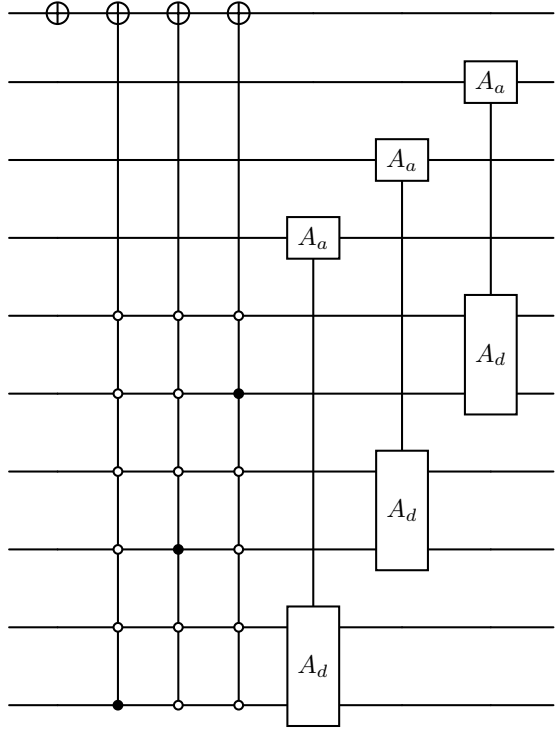
\begin{figure}
\[
\resizebox{0.5\columnwidth}{!}{
\begin{quantikz}
& \targ{} & \targ{} & \targ{} & \targ{} & & & & \\
& & & & & & & \gate{A_a} & \\
& & & & & & \gate{A_a} & & \\
& & & & & \gate{A_a} & & & \\
& & \ctrl[open]{-4} & \ctrl[open]{-4} & \ctrl[open]{-4} & & & \gate[2]{A_d}\wire[u][3]{q} & \\
& & \ctrl[open]{-1} & \ctrl[open]{-1} & \ctrl{-1} & & & & \\
& & \ctrl[open]{-1} & \ctrl[open]{-1} & \ctrl[open]{-1} & & \gate[2]{A_d}\wire[u][4]{q} & & \\
& & \ctrl[open]{-1} & \ctrl{-1} & \ctrl[open]{-1} & & & & \\
& & \ctrl[open]{-1} & \ctrl[open]{-1} & \ctrl[open]{-1} & \gate[2]{A_d}\wire[u][5]{q} & & & \\
& & \ctrl{-1} & \ctrl[open]{-1} & \ctrl[open]{-1} & & & &
\end{quantikz}}
\]
\caption{Block-encoding circuit for $c$. Here, $A_a$ is the $1$-qubit persistent ancilla register of an encoding of $A$ and $A_d$ is the $2$-qubit data register of an encoding of $A$. The sequence of multi-control $X$ gates can be decomposed via unary iteration to reduce their $T$-count to $32$ (to optimally cancel Toffoli gates via Lemma \ref{unaryLEMMA}, expand each multi-controlled gate by operating on the open controls first).}
\label{c}
\end{figure}

\begin{proposition}
The circuit in Figure \ref{c} is a $(0.6495,(5,4),0)$ block-encoding of $c$ costing 44 T-gates. The controlled circuit costs 84 T-gates.
\end{proposition}

In the circuit in Figure \ref{c}, we move all columns of $A^{\otimes 3}$ except those corresponding to $c_x,c_y,c_z$ to an ancilla register. It is not necessary that these columns occupy the first three columns of the unitary since the product $cc^T$ is invariant column permutation in $c$. Since this matrix only involves tensor products and permutations of $A$, the encoding likewise has zero error.

To create the block encoding of $C=cc^T$, we can encode $c^T$ by reversing the circuit in Figure \ref{CFIG}. The product can be encoded by placing these circuits in sequence, and situating the ancilla on separate registers. Control portions are repeated in sequence and thus they only need to be implemented once.

\begin{proposition}
\label{Cprop}
The circuit in Figure \ref{CFIG} is a $(0.4219,(5,8),0)$ block-encoding of $C$ costing $56$ T-gates. The controlled circuit costs $132$ T-gates.
\end{proposition}

\begin{figure}[!ht]
\[
\resizebox{0.6\columnwidth}{!}{
\begin{quantikz}
& & & & \targ{} & \targ{} & \targ{} & \targ{} & & & & \\
& & & \gate{A_a^\dag} & & & & & & & & \\
& & \gate{A_a^\dag} & & & & & & & & & \\
& \gate{A_a^\dag} & & & & & & & & & & \\
& & & & \targ{} & \targ{} & \targ{} & \targ{} & & & & \\
& & & & & & & & & & \gate{A_a} & \\
& & & & & & & & & \gate{A_a} & & \\
& & & & & & & & \gate{A_a} & & & \\
& & & \gate[2]{A_d^\dag}\wire[u][7]{q} & & \ctrl[open]{-8} & \ctrl[open]{-8} & \ctrl{-8} & & & \gate[2]{A_d}\wire[u][3]{q} & \\
& & & & & \ctrl[open]{-1} & \ctrl[open]{-1} & \ctrl[open]{-1} & & & & \\
& & \gate[2]{A_d^\dag}\wire[u][8]{q} & & & \ctrl[open]{-1} & \ctrl{-1} & \ctrl[open]{-1} & & \gate[2]{A_d}\wire[u][4]{q} & & \\
& & & & & \ctrl[open]{-1} & \ctrl[open]{-1} & \ctrl[open]{-1} & & & & \\
& \gate[2]{A_d^\dag}\wire[u][9]{q} & & & & \ctrl{-1} & \ctrl[open]{-1} & \ctrl[open]{-1} & \gate[2]{A_d}\wire[u][5]{q} & & & \\
& & & & & \ctrl[open]{-1} & \ctrl[open]{-1} & \ctrl[open]{-1} & & & &
\end{quantikz}}
\]
\caption{Block-encoding circuit for $C$}
\label{CFIG}
\end{figure}

Once again this block-encoding contains no error. As the $C$ matrix has maximum absolute value $3$, then the matrix $\tilde{C}$ which is being block-encoded by Proposition \ref{Cprop} is given by $\lVert\tilde{C}\rVert =\lVert C\rVert/64$.

We next address a block-encoding of $W$. If 

\begin{equation}
    R_y(\theta _0)=\frac{1}{4}\begin{pmatrix}1 & -\sqrt{15} \\ \sqrt{15} & 1\end{pmatrix},
\end{equation}
then we can easily block-encode $w$ by a single controlled rotation in a $3$-qubit circuit with open control on the second qubit and target on the ancilla. Thus, $W=w^{\otimes 3}$ can be encoded with just $3$ controlled rotation gates.

\begin{figure}
\[
\resizebox{0.35\columnwidth}{!}{
\begin{quantikz}
& & & \gate{R_y(\theta_0)} & \\
& & \gate{R_y(\theta_0)} & & \\
& \gate{R_y(\theta_0)} & & & \\
& & & \ctrl[open]{-3} & \\
& & & & \\
& & \ctrl[open]{-4} & & \\
& & & & \\
& \ctrl[open]{-5} & & &
\end{quantikz}}
\]
\caption{Block-encoding circuit for $W$}
\label{Wfig}
\end{figure}
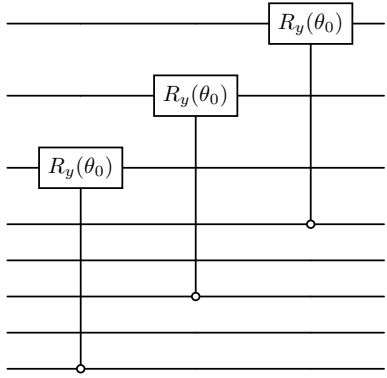

\begin{proposition}
The circuit in Figure \ref{Wfig} is a $(1,(0,3),\epsilon )$ block-encoding of $W$ costing $6.9\log _2(1/\epsilon )+62.1$ T-gates. A controlled application of $W$ costs $6.9\log _2(1/\epsilon )+74.1$ T-gates.
\end{proposition}

{\bf Proof:} If each rotation gate is approximated by $\epsilon$, then this error only affects the factors of $1/4$ in each $w$. Let $y$ be the encoded factor approximating $1/4$ such that $\lvert y-\frac{1}{4}\rvert <\epsilon$. Then it follows that $\lvert y^2-\frac{1}{16}\rvert <\frac{1}{2}\epsilon +\epsilon ^2$ and $\lvert y^2-\frac{1}{16}\rvert <\frac{3}{16}\epsilon +\frac{3}{4}\epsilon ^2+\epsilon ^3$. Since $W$ is a diagonal matrix, the $L^2$ error is given by the maximum of these error quantities, which is simply $\epsilon$. Thus, if the rotation gates are approximated by $\epsilon$ so too is the total $L^2$ error of $W$. Therefore the total $T$ count is given by $3(2.3\log _2(1/\epsilon )+20.7)$. $\hfill\Box$

By a similar yet simplified argument, we can block-encode the matrix $P$ (which is used to zero out any basis states with a $3$ in its base-$4$ representation) with the following circuit:

\begin{proposition}
There exists a circuit that is a $(1,(1,3),0)$ block-encoding of $P$ costing $12$ T-gates.
\end{proposition}

The final piece of equation (\ref{F1}) left to encode is ${\bf 1}_6$. Rather than block-encoding this directly, we instead use the equation ${\bf 1}_6=32(G_6+I_6)$ where $G_6$ is the $6$-qubit Grover matrix with the following resource cost:

\begin{proposition}
There exists a circuit that is a $(1,(4,0),0)$ block-encoding of $G_6$ costing $16$ T-gates and $44$ T-gates controlled.
\end{proposition}

Now that all of the components of equation (\ref{F1}) are accounted for, it remains to construct the LCU to block-encode all of $F_1$. We can rewrite $F_1$ in the suggestive form

\begin{equation}
\label{gateF1}
    F_1=P\left(32W(G_6+I_6+6\tilde{C})-I\right) P.
\end{equation}

This lends itself to a three-ancilla LCU which block-encodes the contained sum as seen in the following lemma:

\begin{lemma}
Let us define the rotation matrices
\begin{equation}
R_y(\theta _1)=\frac{1}{2}\begin{pmatrix} 1 & -\sqrt{3} \\ \sqrt{3} & 1\end{pmatrix},\hspace{1cm}R_y(\theta _2)=\frac{1}{\sqrt{257}}\begin{pmatrix} 16 & -1 \\ 1 & 16\end{pmatrix}
\end{equation}
and the diagonal matrix $D=\text{diag}\left( [a,b,c,c,d,d,d,d]\right)$. Then the matrix
\begin{equation}
    \left(R_y(\theta _2)\otimes R_y(-\theta _1)\otimes H\right) D\left(R_y(\theta _2)\otimes R_y(\theta _1)\otimes H\right)
\end{equation}
has its top left entry equal to
\begin{equation}
    \frac{1}{257}\left(32(a+b+6c)-d\right).
\end{equation}
\end{lemma}

This lemma gives us a blueprint for creating the PREP oracles going into the LCU. We can now present the main result.

\begin{theorem}
\label{gateTHM}
There exists a circuit that is a $(0.0225,(7,20),\epsilon )$ block-encoding of $F_1$ costing \\ $11.5\log_2\left( 1/\epsilon\right) + 404.5$ $T$-gates.
\end{theorem}
{\bf Proof:} See Appendix \ref{circuits}.

While the subnormalization for the circuit in Theorem \ref{gateTHM} is about four times larger than the subnormalizations resulting from the unstructured oracle methods, this implementation of the $F_1$ block-encoding has a far lower $T$-cost. The ratio of $T$-cost to subnormalization at at an error rate of $\epsilon =10^{-10}$ indicates that the algebraic method is about $3.2$ times more efficient than the QROM implementation, $22$ times more efficient than unary iteration, and $77$ times more efficient than S-FABLE.

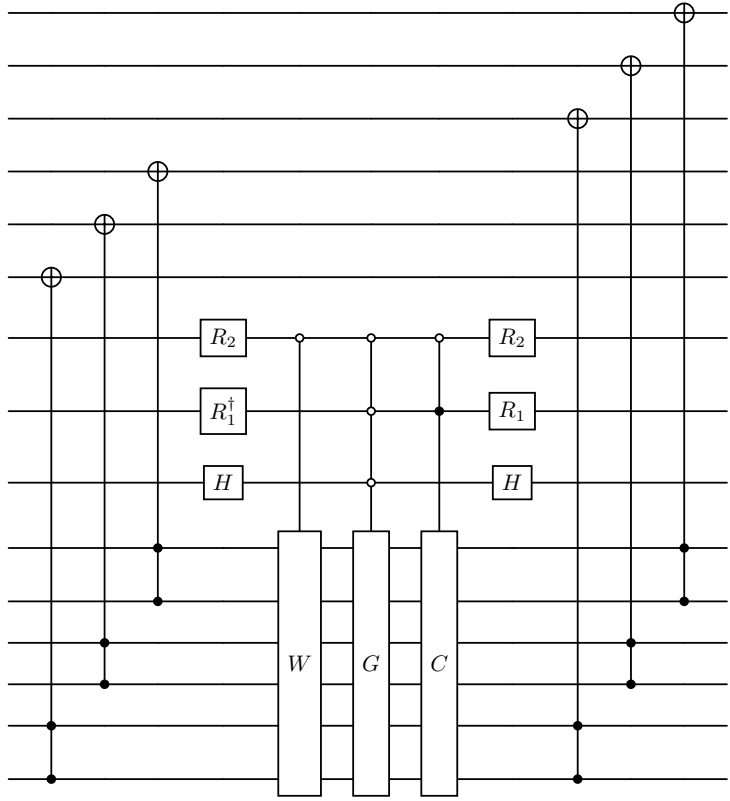
\begin{figure}
\[
\resizebox{0.65\columnwidth}{!}{
\begin{quantikz}
& & & & & & & & & & & \targ{} & \\
& & & & & & & & & & \targ{} & & \\
& & & & & & & & & \targ{} & & & \\
& & & \targ{} & & & & & & & & & \\
& & \targ{} & & & & & & & & & & \\
& \targ{} & & & & & & & & & & & \\
& & & & \gate{R_2} & \ctrl[open]{3} & \ctrl[open]{1} & \ctrl[open]{1} & \gate{R_2} & & & & \\
& & & & \gate{R_1^\dag} & & \ctrl[open]{1} & \ctrl{2} & \gate{R_1} & & & & \\
& & & & \gate{H} & & \ctrl[open]{1} & & \gate{H} & & & & \\
& & & \ctrl{-6} & & \gate[6]{W} & \gate[6]{G} & \gate[6]{C} & & & & \ctrl{-9} & \\
& & & \ctrl{-1} & & & & & & & & \ctrl{-1} & \\
& & \ctrl{-7} & & & & & & & & \ctrl{-10} & & \\
& & \ctrl{-1} & & & & & & & & \ctrl{-1} & & \\
& \ctrl{-8} & & & & & & & & \ctrl{-11} & & & \\
& \ctrl{-1} & & & & & & & & \ctrl{-1} & & &
\end{quantikz}}
\]
\caption{Block-encoding circuit for $F_1$}
\label{F1fig}
\end{figure}

\paragraph{Subnormalization optimized}

While the previous method drastically reduces $T$-count compared to the unstructured oracle encodings, the subnormalization is about four times more costly. As we will see in the following section, this will carry ramifications if there are many additional components in the circuit which are also subject to the subnormalization of the block-encoding. 

\begin{figure}
\[
\begin{quantikz}
& \gate{R_y(\theta)} & \ctrl[open]{1} & \\
& & \gate{H} &
\end{quantikz}
\]
\caption{Block-encoding circuit for $A$ with unit subnormalization}
\label{A2}
\end{figure}
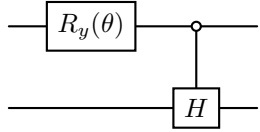

Many factors contribute to this tiny subnormalization, but one that can simply be addressed occurs at the very beginning; the block-encoding of $A$ carries a subnormalization of $\sqrt{3}/2$. While this does not sound particularly egregious, in the block-encoding of $C$ in Figure \ref{CFIG} we call this circuit $6$ times, thus the total contribution to the subnormalization of this piece is $(\sqrt{3}/2)^6\approx 0.42$. If we could improve the subnormalization of $A$ to be $1$, then this could lead to a nearly $2.4\times$ improvement in subnormalization; this is possible per the following proposition:

\begin{proposition}
Let us define the rotation matrix 
\begin{equation}
    R_y(\theta )=\frac{1}{\sqrt{3}}\begin{pmatrix}\sqrt{2} & -1 \\
    1 & \sqrt{2} \end{pmatrix}.
\end{equation} Then the circuit in Figure \ref{A2} is a $(1,(0,0),\epsilon )$ block-encoding which costs $1.15\log _2(1/\epsilon )+11.2$ T-gates uncontrolled and $2.3\log _2(1/\epsilon )+26.7$ T-gates controlled.
\end{proposition}

While the subnormalization is ideal in this circuit, it comes at the cost of introducing a rotation gate with a comparatively large $T$-count. As we will find, this sacrifice will be worth it in certain settings. We repeat the steps of the previous procedure to build a block-encoding of $F_1$ using this new block-encoding of $A$. We thus present a `subnormalization optimized' block-encoding of $F_1$ with details laid out in Appendix \ref{circuits}:

\begin{theorem}
\label{subTHM}
There exists a circuit that is  a $(0.0397,(7,14),\epsilon )$ block-encoding of $F_1$ costing \\ $25.3\log _2(1/\epsilon )+597.9$ T-gates.
\end{theorem}
{\bf Proof:} See Appendix \ref{circuits}.

At the cost of six additional controlled rotations and about 200 $T$-gates of approximation overhead, theorem \ref{subTHM} provides a $76\%$ improvement in subnormalization over theorem \ref{gateTHM}.
    \section{Method Comparisons}
\label{four}

\begin{figure}
\centering
\includegraphics[scale=0.18]{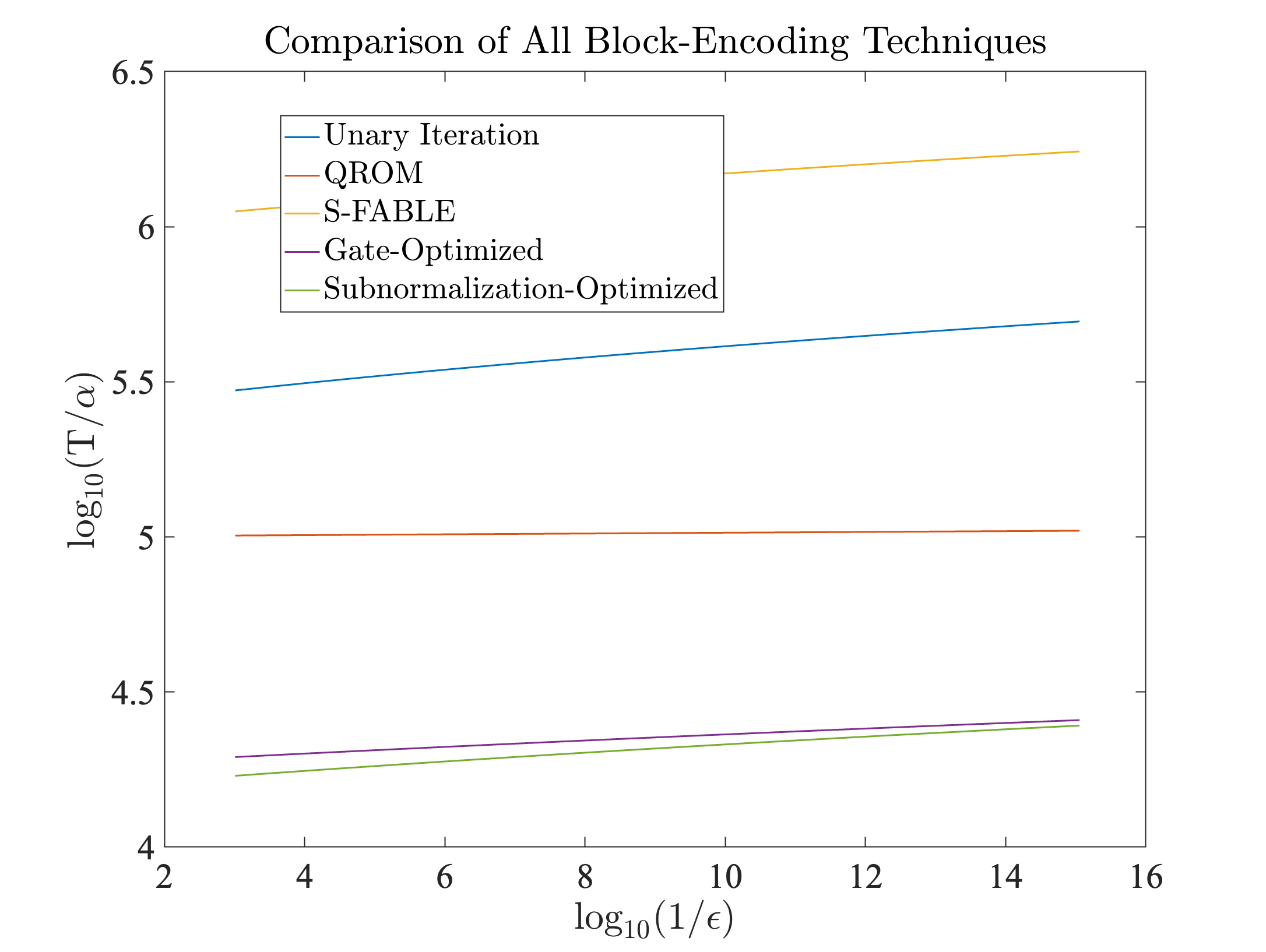}
\caption{Comparison of block-encoding methods measured by the metric of $T$-count divided by subnormalization.}
\label{comparision}
\end{figure}

From the theorems presented in Section \ref{three}, we now compare block-encoding methods for the matrix $F_1$. In Figure \ref{comparision}, we use the ratio of $T$-count to subnormalization $T(\epsilon)/\alpha$ to concretely estimate the efficiency of each method. As expected, there is a clear stratification of the unstructured oracle methods, with QROM being the most efficient and S-FABLE the least. These are seperated by an order of magnitude in cost. Meanwhile, both algebraic methods soundly beat the unstructured methods, with the subnormalization-optimized method being slightly more favorable than gate-optimized when $\epsilon >2.7\times 10^{-19}$; when $\epsilon =10^{-10}$, there is a $3.6\%$ improvement in this ratio for the subnormalization-optimized encoding over the gate-optimized encoding.

This single metric does not capture the whole picture, however. In particular, Figure \ref{comparision} assumes that the block-encoding is the only component of the circuit that has a $T$-cost. To give a different perspective, consider a block-encoding of $F_1$ to apply to a $6$-qubit state $|\psi\rangle$ which costs $x$ $T$-gates to prepare (assume this state-preparation is errorless). In this scenario, the probability of observing $F_1|\psi\rangle$ will scale inversely with the subnormalization, meaning that we need to repeat both the block-encoding and the state-preparation more times in an amplitude amplification. Thus, if a block-encoding method can be implimented using $a\log _2(1/\epsilon )+b$ $T$-gates and has subnormalization $\alpha$, it is helpful to define a new quantity of interest $(a\log _2(1/\epsilon )+b+x)/\alpha$. Since $\epsilon$ and $x$ are variables, we can separate the $(x,\log _2(1/\epsilon))$ plane into regions which favor certain block-encoding methods. When comparing the three best performing methods, QROM is advantageous when $x$ is large, conversely the gate-optimized encoding is best for small $x$ and relatively high accuracy constraints. In practice block-encodings are unlikely to require approximation errors below $\epsilon <10^{-20}$ and the state-preparation circuit for a $6$-qubit state is likewise unlikely to require more than a few thousand $T$-gates (arbitrary state-preparation would cost at worst $2^6$ controlled rotations). Stated differently, the subnormalization-optimized encoding is superior for realistic problems.

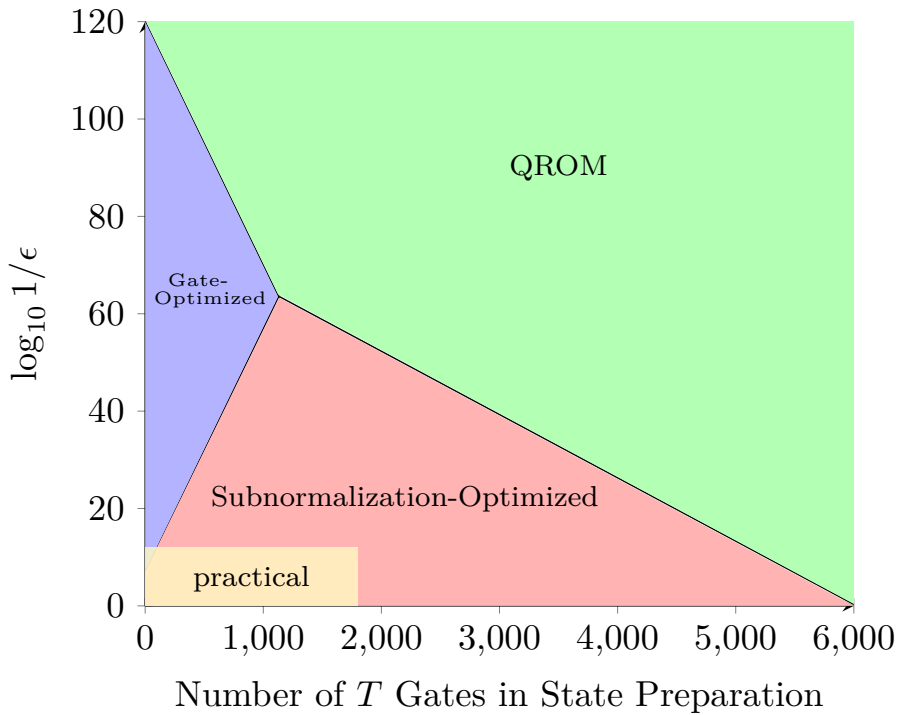
\begin{figure}
\centering
\resizebox{350pt}{!}{
\begin{tikzpicture}
\begin{axis}[
    axis lines = left,
    ymin = 0,
    xlabel = Number of $T$ Gates in State Preparation,
    ylabel = {\(\log_{10}{1/\epsilon}\)},
]
\addplot [
    domain=0:42000/37,
    samples=1000, 
    color=black,
    name path=A,
]
{-0.05*x + 120};
\addplot [
    domain=42000/37:6000,
    samples=1000, 
    color=black,
    name path=B,
]
{-0.013*x + 78.3};
\addplot [
    domain=0:42000/37,
    samples=1000, 
    color=black,
    name path=C,
]
{0.05*x + 7};
\fill [blue!30]
    (42000/37-10,63.5) -- (0,7.25) -- (0,120) -- cycle;
\fill [green!30]
    (42000/37-10,64) -- (0,120.5) -- (0,125) -- (6030,125) -- (6030,0) -- cycle;
\fill [red!30]
    (42000/37,63.3) -- (6000,0) -- (0,0) -- (0,7.1) -- cycle;
\fill[yellow!30,opacity=0.8] (0,0) rectangle (1800,12);
\node at (900,5.5) {\footnotesize practical};
\node at (2200,22)  {\footnotesize Subnormalization-Optimized};
\node at (450,67) {\tiny Gate-};
\node at (550,63) {\tiny Optimized};
\node at (3500,90) {\footnotesize QROM};

\end{axis}
\end{tikzpicture}
}
\caption{Plot of advantageous regimes for the state-preparation problem for each block-encoding method. Since the regimes for which gate-optimized and QROM strategies are unrealistic, the subnormalization-optimized method is ideal.}
\end{figure}

\subsection{Average Error Estimates}

Now that we have given bounds on the actual number of $T$-gates required to block-encode $F_1$ to accuracy $\epsilon$ in section \ref{three} we will pursue a more nuanced form of resource estimation. In practical fault-tolerant settings, the rotation gates will be approximated using series of one-qubit transformations with explicit $T$-counts \cite{bocharov15}. We have focused our discussion on a matrix $F_1$ that is small enough to simulate the actual $L^2$ error. While in the proofs we place upper bounds on the error by assuming all individual error vectors are pointing in the same direction, we can significantly undershoot these bounds by choosing rotation approximations with error vectors pointing in different directions.

To ease the cost of simulation, rather than computing a unique gate sequence for each rotation $R_y(\theta )$, we find a matrix $U$ such that $\lVert R_y(\theta )-U\rVert\approx\epsilon$ and assume it costs exactly $1.15\log _2(1/\epsilon )+9.2$ to implement $U$. To construct $U$, we use $R_z(\pm\epsilon \sqrt{\sigma})R_x(\pm\epsilon\sqrt{1-\sigma})R_y(\theta )$ where $\sigma\sim U(0,1)$. Rather than using different approximations for each of the constituent uncontrolled rotations as in Figure \ref{controlledRotation}, we will simplify the analysis by finding one approximation for the first and implementing its inverse for the second; for a controlled $y$ rotation the approximation this takes is:

\begin{equation}
    \begin{pmatrix} I & 0 \\
    0 & U\end{pmatrix}.
\end{equation} 

Rotations from our unstructured oracle circuits are approximated to a constant $\epsilon$. In the case of the algebraic methods the rotations are approximated according to the ratios calculated by the Lagrangian multiplier optimization detailed in Appendix \ref{circuits}.

We summarize the differences between the upper bound and average costs in Table \ref{table}. These counts underscore that the upper bounds are not too loose compared to the average case; overheads improve roughly $20\%$ with most methods. The average QROM cases only offer about $200$ $T$-gates of improvement over the upper bound while an average unary iteration case is about $4500$ $T$-gates cheaper. This is because when performing unary iteration we can randomize the rotation approximations for each individual element in the matrix. We can visualize this by noting that error vectors will point in random directions on the Bloch sphere, leading to more cancellation and a smaller total $L^2$ error. Errors for each unique element will point in the same direction for QROM methods which is comparable with the worst-case scenario of errors all pointing in the same direction.

\begin{table}
\centering
\begin{tabular}{ |c|c|c| } 
 \hline
 {\bf Block-encoding method} & {\bf Upper bound} & {\bf Average error estimate} \\ 
 \hline
 Unary Iteration & $1490.4\log _2(1/\epsilon )+22368$ & $1491.9\log _2(1/\epsilon )+17950$ \\ 
 QROM & $27.6\log _2(1/\epsilon )+9062$ & $27.8\log _2(1/\epsilon )+8888$ \\ 
 S-FABLE & $4710.4\log _2(1/\epsilon )+86999$ & $4753.8\log _2(1/\epsilon )+73688$ \\ 
 Gate-optimized & $11.5\log _2(1/\epsilon )+404.5$ & $11.5\log _2(1/\epsilon )+352.6$ \\ 
 Subnormalization-optimized & $25.3\log _2(1/\epsilon )+597.9$ & $25.2\log _2(1/\epsilon )+500.9$ \\ 
 \hline
\end{tabular}
\caption{Comparison of T-counts between upper bounds presented in Section \ref{three} and average counts simulated by the method described in Section \ref{four}.}
\label{table}
\end{table}

One could hypothetically choose rotation approximations to optimize the $T$-count. We do not present those results here as our rotation gate approximations are purely synthetic (i.e. not created from actual sequences of Clifford+$T$). However, simulated data suggests that overheads could be reduced as much as $35\%$ from the upper bound. We also mention that the algebraic circuits can be significantly parallelized to optimize $T$-depth, while the unstructured circuits are inherently serial.
    \section{Conclusion}

We have constructed a variety of block-encoding circuits for an application-derived six-qubit matrix and compared their efficiency. This matrix was chosen because it has many advantageous properties for an unstructured encoding, i.e. it is small, has many repeated entries, and is sparse. Even when minding problem structure, we could only generate block encodings that are fairly complex and have poor subnormalization. Nonetheless, these structured encodings performed close to an order of magnitude better than the best unstructured encodings. We expect this discrepancy to become even more prominent for larger matrices as the $T$-costs associated with unstructured techniques scale exponentially. An important finding through this design process is that although FABLE was introduced as a promising alternative to a standard unary iteration since it completely removed multi-controls this is offset by the massive rotation approximation overheads, and should not be used in a fault-tolerant setting. Based on these lessons, the authors urge the community to explore block-encoding techniques beyond complexity theory results. Generic methods are often riddled with hidden costs, while there is a vast array of under-studied, clever, bespoke matrix-algebraic methods which are more natural for block encoding matrices.
    
    \section*{Acknowledgments}

    All authors acknowledge support from the Defense Advanced Research Projects Agency under Air Force Contract No. FA8702-15-D-0001. Any opinions, findings and conclusions or recommendations expressed in this material are those of the authors and do not necessarily reflect the views of the Defense Advanced Research Projects Agency.
    
    \small
    \bibliographystyle{plainnat}
    \bibliography{main.bib}
    \normalsize
    
    \begin{appendices}
    \section{Circuits Related to Section \ref{three}}
\label{circuits}
\subsection{Unstructured}

\textbf{Proof of Theorem \ref{unaryTHM}:} \\
Since $\text{max}(F_1)=1$, it is simple enough to calculate the $L^2$ subnormalization factor for the unstructured oracle implementation as $\lVert F_1\rVert /64\approx 0.0905$.

The number of ancilla falls right out of the unstructured oracle construction; $2n-1$ clean ancilla and $n+1$ persistent ancilla.

Contributing to the total $T$-count of this circuit is the multi-control sequence and the controlled rotations themselves. By using Lemma \ref{unaryLEMMA}, we can cut down the number of Toffoli pairs in the control sequence from 7942 ($722$ nonzero elements times $11$ Toffoli pairs per multi-controlled rotation) down to $1085$ or $4340$ $T$-gates.

Of the $722$ nonzero elements, $648$ contain something other than a maximal element, so we only need to implement $648$ controlled rotations. Suppose errors are distributed evenly such that each rotation is approximated by $\epsilon$. Then if $\tilde{a}_ij$ is the actual entry encoded in the block from these approximate rotations, we have $|a_{ij}-\tilde{a}_{ij}|<\epsilon /64$ for all $i,j$. We can create an upper bound on the total error by directing all rotation errors in the same `direction', i.e. $A-\tilde{A}$ is a matrix with $\epsilon /64$ in place of each element that isn't a zero or a one. From the particular structure of $F_1$, we can place the norm of this error as $\lVert A-\tilde{A}\rVert\le 0.3867\epsilon$. Since $A=F_1/64$ such that $\lVert A\rVert =0.0905$, by Lemma \ref{lemmaBOUND} we can bound the normalized error as $\lVert A/\lVert A\rVert -\tilde{A}/\lVert\tilde{A}\rVert\rVert\le 8.55\epsilon$. Therefore, if each rotation is approximated by $\epsilon /8.55$ then the total block-encoding has an $L^2$ error bounded by $\epsilon$. Thus, combining these figures into an upper bound on the $T$-cost of this block-encoding gives us
$4340+648\left( 2.3\log _2(8.55/\epsilon )+20.7\right)$ which simplifies to the stated result. $\hfill\Box$ \\

\noindent
\textbf{Proof of Theorem \ref{QROMTHM}:} \\
The subnormalization is the same as in the unary iteration implementation since the QROM circuit is just an alternative construction of the unstructured oracle.

We use an extra $4$ clean ancilla as a $4$ qubit register to store the unique rotation angles. 

The QROM oracle can be expressed as two separate oracles, the first mapping $|i\rangle |j\rangle |0\rangle |0\rangle\mapsto|i\rangle |j\rangle |u_{ij}\rangle |0\rangle$ where $u_{ij}\in\{ 0,...,13\}$ denotes which of the $14$ unique elements will be encoded in entry $(i,j)$. The second acts on the last two registers as $|n\rangle |0\rangle\mapsto |n\rangle\left( R_y(\theta _n)|0\rangle\right)$; we say $\theta _0=0$ so that we can still use sparsity to simplify the construction. Finally we invert the first oracle to uncompute the $4$-qubit `unique element' register.

Since we can write any function $|0\rangle\mapsto |n\rangle$ with only $X$ gates, the first oracle accumulates $T$-gates only from the control pattern which is the same as in Theorem \ref{unaryTHM}. As we use both this oracle and its inverse the total contribution is $8680$ $T$-gates.

The second oracle consists of $12$ controlled rotations and a multi-controlled $X$ gate. Using Lemma \ref{unaryLEMMA} this control sequence costs $12$ Toffoli pairs or $48$ $T$-gates. Since we can use the same error analysis from the proof of \ref{unaryTHM}, if these rotations have error $\epsilon$, then the total error on the block-encoding is given by $8628+12(2.3\log _2(8.55/\epsilon )+20.7)$ which simplifies to the stated result.
$\strut\hfill\Box$

\noindent
\textbf{Proof of Theorem \ref{FABLE_THM}:} \\
Since the S-FABLE method uses FABLE to encode $H^{\otimes n}F_1H^{\otimes n}$ and $\lVert H^{\otimes n}F_1H^{\otimes n}\rVert _\infty =1.002$, this slightly reduces the subnormalization compared to a straightforward oracle encoding of $F_1$.

Since the S-FABLE oracle consists only of single qubit rotations and CNOT gates there are no clean ancilla necessary, only seven persistent ancilla.

To compute the angles for the rotation gates, let $\text{vec}(\theta )$ be an initial set of angles such that 
\begin{equation}
    \theta _{ij}=\cos ^{-1}\left( \frac{(H^{\otimes n}F_1H^{\otimes n})_{ij}}{\lVert H^{\otimes n}F_1H^{\otimes n}\rVert _\infty}\right).
\end{equation}
The set of angles $\hat{\theta}$ implemented in the rotation gates satisfy $\text{vec}(\hat{\theta})=P_G^{-1}H^{\otimes 12}\text{vec}(\theta )/64$ where $P_G$ is a permutation carrying integers to a Gray code ordering \cite{doran07}.

To take advantage of sparsity in the S-FABLE construction, we bound the magnitudes of the most significant rotations in $\text{vec}(\hat{\theta})$ by a simple function; for this purpose let us order these angles such that $|\hat{\theta}_i|\ge |\hat{\theta}_{i+1}|$. We consider the largest 664 rotations to be the most significant such that $|\hat{\theta}_i|>5.5\times 10^{-4}$ for $i\le 628$. For the remainder, we calculate that $|\hat{\theta}_i|<10^{-0.00039i-3}$. This leads to the existence of three regimes. Suppose we approximate each rotation by some $\epsilon$; if $\epsilon >5.5\times 10^{-4}$ we can eliminate all but the most relevant $628$ rotations in the circuit, if $\epsilon <4.99\times 10^{-7}$ then we keep all $4096$ rotations, and if $\epsilon$ takes an intermediate value between these quantities we will keep $\lceil 2564\log _{10}(1/\epsilon)-7692\rceil$ rotations. Call this piecewise function $k(\epsilon )$.

\begin{figure}[!ht]
\centering
\includegraphics[scale=0.18]{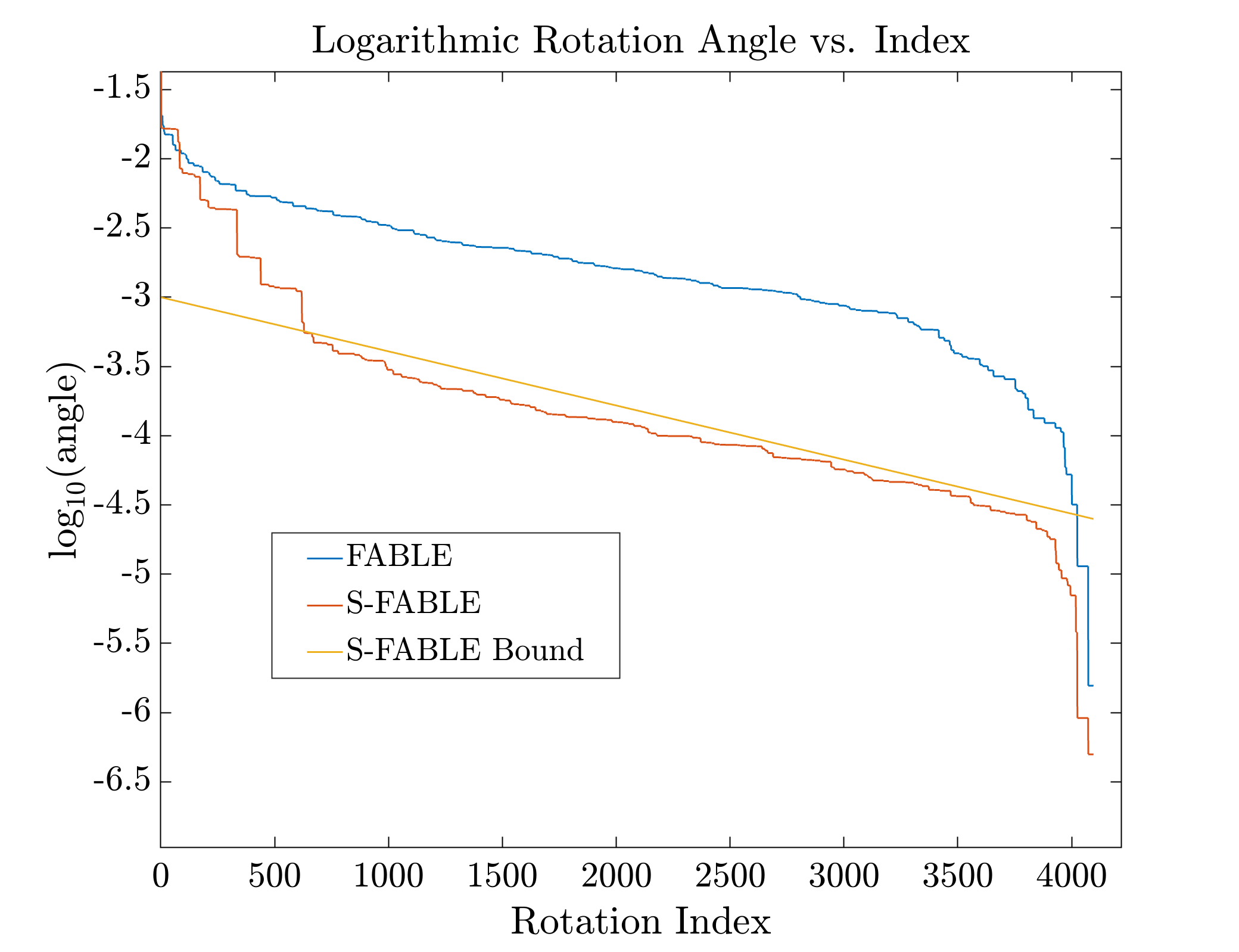}
\caption{Plot of rotation angles in FABLE and S-FABLE circuits, ordered with decreasing magnitude. Yellow curve is exponential in the index and bounds the S-FABLE rotations past index $664$. Notice that the S-FABLE angles are not fully eliminated as in a sparse matrix, rather they are suppressed compared to the less relevant angles of FABLE.}
\end{figure}

Regardless of the value of $k(\epsilon )$, if we approximate each rotation angle up to $\epsilon$ then we can produce a coarse bound on the total circuit error by assuming these errors are pointing in the same direction in $\mathbb{C}$; i.e. if $\text{vec}(\hat{\theta}(\epsilon ))$ is the vector of approximate rotations then consider $\text{vec}(\hat{\theta})-\text{vec}(\hat{\theta}(\epsilon ))=\epsilon {\bf 1}_{4096\times 1}$. Propagating this through the FABLE equations to a block-encoding, if $\tilde{F}_1$ is the approximated matrix by taking cosines of the approximated initial angle set then it is easy to show $\lVert F_1-\tilde{F}_1\rVert <4096\epsilon$ ($F_1-\tilde{F}_1$ is empty except for $4096\epsilon$ in the top left corner). By Lemma \ref{lemmaBOUND}, the normalized error on the total block-encoding can be bound by $1418\epsilon$. Therefore if we approximate each rotation by $\epsilon /1418$, the total block-encoding error will be $\epsilon$ and the total $T$-count of the circuit becomes $k\left(\frac{\epsilon}{1418}\right)(1.15\log _2\left(\frac{1418}{\epsilon}\right)+9.2)$ which simplifies to the stated result. Since the top regime bound becomes transformed to $1418(5.5\times 10^{-4})=0.78$, we neglect to report the first regime. $\hfill\Box$
    \subsection{Algebraic}

\subsubsection{Gate Optimized}
One final piece not addressed in the main text is the total error in an LCU circuit provided that both the PREP and SELECT elements have errors themselves. The following lemma treats this issue for the two single-qubit LCU circuits which will be contained in our larger LCU:
\begin{lemma}
\label{bb1}
Let us define quantities $p_1,p_2,\tilde{p}_1,\tilde{p}_2\in\mathbb{C}$ and operators $A,B,C,\tilde{C}$ such that $|p_1|^2+|p_2|^2=1$, $|p_1-\tilde{p}_1|^2+|p_2-\tilde{p}_2|^2<\epsilon _0^2$ and $\lVert C-\tilde{C}\rVert <\epsilon _1$. Then 
\begin{align}
\label{error0}
    \left\lVert\left( |p_1|^2CA+|p_2|^2CB\right) -\left( |\tilde{p}_1|^2\tilde{C}A+|\tilde{p}_2|^2\tilde{C}B\right)\right\rVert &<\left( 2\lVert C\rVert\epsilon _0+\epsilon _1\right)\sqrt{\lVert A\rVert ^2 +\lVert B\rVert ^2} \\
\label{error1}
    \left\lVert\left( |p_1|^2C-|p_2|^2A\right) -\left( |\tilde{p}_1|^2\tilde{C}-|\tilde{p}_2|^2A\right)\right\rVert &<2\epsilon_0\sqrt{\lVert A\rVert ^2+\lVert C\rVert ^2}+\epsilon _1
\end{align}
\end{lemma}
{\bf Proof:} The result follows from using the approximation $|xy-\tilde{x}\tilde{y}|<|y||x-\tilde{x}|+|x||y-\tilde{y}|$ (suitable approximation since the missing term incorporating the product of the errors is exceptionally small) and applying Cauchy-Schwarz. $\hfill\Box$ \\

\noindent
\textbf{Proof of Theorem \ref{gateTHM}:} \\
The circuit in Figure \ref{F1fig} block-encodes $F_1/257$ when $A$ is given by Figure \ref{gateA} which gives us the desired subnormalization constant.

Counting the persistent ancilla, eight come from $C$, another three from $W$, three come from implementing the LCU, and six come from the Toffoli gates implementing the $P$ matrices, leading to seventeen in total. Eight clean qubits arise from the multi-controlled gates in $C$ which are controlled on nine qubits.

We now count the $T$-gates in our circuit construction. Considering the three sources of approximation in the circuit, assume $W$ is approximated to $\epsilon _0$, each $R_y(\theta _1)$ is approximated to $\epsilon _1$, and $R_y(\theta _2)$ is approximated to $\epsilon _2$. Recognizing the multi-controls on $G$ and $C$ cost $8$ $T$-gates to implement, the total circuit has $T$-cost
\begin{equation}
    6.9\log_2\left(\frac{1}{\epsilon _0}\right) +2.3\log_2\left(\frac{1}{\epsilon _1}\right) +2.3\log_2\left(\frac{1}{\epsilon _2}\right) +318.9
\end{equation}

We now compute the total error of the circuit. First consider the sum over $R_y(\theta _1)$ which implements $\frac{1}{64}WP\left(\frac{1}{4}{\bf 1}_6+\frac{3}{4}C\right) P$ in the top-left block (we address conjugation by $P$ at this step rather than at the end). Since $W$ has error $\epsilon _0$, $R_y(\theta _1)$ has error $\epsilon _1$, and the other components carry no error, it follows from the equation (\ref{error0}) that the unnormalized error from this component is $0.507(\epsilon _0+2\epsilon _1)$. Plugging this quantity into equation (\ref{error1}) gives us an unnormalized error on the block-encoding of $F_1$ of approximately $\frac{1}{2}\epsilon _0+\epsilon _1+2\epsilon _2$. Since the matrix encoded is an approximation of $F_1/257$, we can use Lemma \ref{lemmaBOUND} to place the total normalized error at 
\begin{equation}
\label{merit0}
    44.39(\epsilon _0+2\epsilon _1+4\epsilon _2).
\end{equation}

Now that we have a $T$-count and a total circuit error in terms of the individual error quantities, we can minimize $T$-count subject to an error constraint $\epsilon$. Using the method of Lagrangian multipliers, we can calculate that the $T$-count is maximized for a circuit with total error $\epsilon$ when $\epsilon _0=12\epsilon /\lambda$, $\epsilon _1=2\epsilon /\lambda$, and $\epsilon _2=\epsilon /\lambda$ where $\lambda\approx 887.8$. Substituting these values into equation (\ref{merit0}) gives us the final result. $\hfill\Box$

\subsubsection{Subnormalization Optimized}

\begin{proposition}
\label{C2}
There exists a circuit that is a $(1,(5,2),\epsilon )$ block-encoding of $C$ which costs \\ $6.9 \log _2(1/\epsilon ) + 117$ T-gates uncontrolled and $13.8\log _2(1/\epsilon )+231.9$ T-gates controlled.
\end{proposition}
{\bf Proof:} Result follows from recognizing that if $A$ is approximated with error $\epsilon$, then the total circuit error is bounded by $6\epsilon$. $\hfill\Box$

Recognize that the above block-encoding is now strictly encoding $\tilde{C}=C/27$. With this new representation, we must alter equation (\ref{gateF1}) to
\begin{equation}
    F_1=P\left( W(32(G_6+I_6)+81\tilde{C})-I\right) P.
\end{equation}
This also requires a new collection of rotation gates to be implemented in the LCU to recover a block-encoding of $F_1$:
\begin{lemma}
Let us define the rotation matrices
\begin{equation}
R_y(\theta _1)=\frac{1}{\sqrt{145}}\begin{pmatrix} 8 & -9 \\ 9 & 8\end{pmatrix},\hspace{1cm}R_y(\theta _2)=\frac{1}{\sqrt{146}}\begin{pmatrix} \sqrt{145} & -1 \\ 1 & \sqrt{145}\end{pmatrix}
\end{equation}
and the diagonal matrix $D=\text{diag}\left( [a,b,c,c,d,d,d,d]\right)$. Then the matrix
\begin{equation}
    \left(R_y(\theta _2)\otimes R_y(-\theta _1)\otimes H\right) D\left(R_y(\theta _2)\otimes R_y(\theta _1)\otimes H\right)
\end{equation}
has its top left entry satisfying
\begin{equation}
    \frac{1}{146}\left(32(a+b)+81c-d\right).
\end{equation}
\end{lemma}

We also consider the following generalization of Lemma \ref{bb1} which we will need in the updated LCU now that the block-encoding from $C$ contains an error term:
\begin{lemma}
\label{bb2}
Let us define quantities $p_1,p_2,\tilde{p}_1,\tilde{p}_2\in\mathbb{C}$ and operators $A,B,\tilde{A},\tilde{B}$ such that $|p_1|^2+|p_2|^2=1$, $|p_1-\tilde{p}_1|^2+|p_2-\tilde{p}_2|^2<\epsilon _0^2$ and $\lVert A-\tilde{A}\rVert <\epsilon _A$, and $\lVert B-\tilde{B}\rVert <\epsilon _B$. Then 
\begin{equation}
    \left\lVert\left( |p_1|^2A+|p_2|^2B\right) -\left( |\tilde{p}_1|^2\tilde{A}+|\tilde{p}_2|^2\tilde{B}\right)\right\rVert <2\epsilon _0\sqrt{\lVert A\rVert ^2 +\lVert B\rVert ^2}+\sqrt{\epsilon _A^2 +\epsilon _B^2}
\end{equation}
\end{lemma}

\noindent
\textbf{Proof of Theorem \ref{subTHM}:} \\
The circuit in Figure \ref{F1fig} encodes $F_1/146$ when using $A$ from Figure \ref{A2} leading to the desired subnormalization.

Since the implementation of $C$ from Proposition \ref{C2} only contains two ancilla the persistent ancilla count from theorem \ref{gateTHM} is reduced by six. The clean ancilla retains the same count as in theorem \ref{gateTHM}.

To count the number of gates, we simply replace the $132$ gates from the implementation of $C$ in its gate-optimized circuit with the new implementation. If we approximate $C$ to error $\epsilon _C$, $W$ to error $\epsilon _W$, and the rotations to $\epsilon _1,\epsilon _2$, then the $T$-count is given by
\begin{equation}
\label{TCOUNT}
    13.8\log _2\left(\frac{1}{\epsilon _C}\right) +6.9\log _2\left(\frac{1}{\epsilon _W}\right) +2.3\log _2\left(\frac{1}{\epsilon _1}\right) +2.3\log _2\left(\frac{1}{\epsilon _2}\right) +418.8.
\end{equation}

We now compute the total circuit error. The sum over the first rotation $R_y(\theta _1)$ implements $\frac{64}{145}\left(\frac{1}{64}WP{\bf 1}_6P\right)+\frac{81}{145}\left(\frac{1}{27}WPCP\right)$. Since error arises from both $W$ and $C$, using the now updated version of Lemma \ref{bb2} we can place the unnormalized error at
$$0.23\epsilon _1+\sqrt{(0.422\epsilon _W)^2+(\epsilon _C+\frac{2}{3}\epsilon _W)^2}$$
We bound the square root in this expression by $\epsilon _C+0.789\epsilon _W$ for convenience. Passing this expression through the LCU induced by the second rotation, we can re-use the second equation from Lemma \ref{bb1} to obtain the total unnormalized error. Finally, to pass this through to the normalized error, we use Lemma \ref{lemmaBOUND} to set the final total at
\begin{equation}
\label{errorTOTAL}
    50.43\left( 0.23\epsilon _1+2\epsilon _2+\epsilon _C+0.789\epsilon _W\right).
\end{equation}

We now optimize the $T$-count provided in equation (\ref{TCOUNT}) subject to the constraint that the expression in equation (\ref{errorTOTAL}) equals a constant $\epsilon$. Working through the Lagrange multiplier calculations, we find that the optimal errors occur in a ratio of $(\epsilon _C:\epsilon _W:\epsilon _1:\epsilon _2) =(6:3.8:4.35:0.5)$. If we multiply each quantity in this ratio by $\epsilon /554.66$, then the total error on the circuit is bound by $\epsilon$ and the desired gate count emerges. $\hfill\Box$
    \end{appendices}

\end{document}